\documentclass[conference]{IEEEtran}
\IEEEoverridecommandlockouts

\usepackage{enumitem} 
\usepackage{cite}
\usepackage{amsmath,amssymb,amsfonts}
\usepackage{mathtools}
\usepackage{graphicx}
\usepackage{caption}
\usepackage{textcomp}
\usepackage{xcolor}
\usepackage{comment}
\usepackage{tabularx,ragged2e,booktabs}
\usepackage{makecell}
\usepackage{multirow}
\usepackage[most]{tcolorbox}
\usepackage[justification=centering]{caption}
\usepackage[utf8]{inputenc}
\usepackage{longtable}
\usepackage{caption}
\usepackage{lipsum}
\usepackage{algorithm} 
\usepackage{algpseudocode} 
 \usepackage{float}

\def\BibTeX{{\rm B\kern-.05em{\sc i\kern-.025em b}\kern-.08em
    T\kern-.1667em\lower.7ex\hbox{E}\kern-.125emX}}

\begin{document}

\title{Unraveling Latch Locking Using Machine Learning, Boolean Analysis, and ILP \\
\thanks
{This material is based on research sponsored by the Air Force Research Labs (AFRL) and the Defense Advanced Research Projects Agency (DARPA) under agreement number \#FA8650-18-1-7817. The U.S. Government is authorized to reproduce and distribute reprints for Governmental purposes notwithstanding any copyright notation thereon. The views and conclusions contained herein are those of the authors and should not be interpreted as necessarily representing the official policies or endorsements, either expressed or implied, of the U.S. Government.}

}
\author{
\IEEEauthorblockN{Dake Chen\textsuperscript{1}, Xuan Zhou\textsuperscript{1}, Yinghua Hu\textsuperscript{1}, Yuke Zhang\textsuperscript{1}, Kaixin Yang\textsuperscript{1}, Andrew Rittenbach\textsuperscript{2}, \\ Pierluigi Nuzzo\textsuperscript{1} and Peter A. Beerel\textsuperscript{1}}
\IEEEauthorblockN{\textsuperscript{1}University of Southern California, \textsuperscript{2}USC Information Sciences Institute} 
\IEEEauthorblockA{\{dakechen, zhouxuan, yinghuah, yukezhan, kaixinya, nuzzo, pabeerel\}@usc.edu, arittenb@isi.edu}\\}

\maketitle

\begin{abstract}
Logic locking has become a promising approach to provide hardware security in the face of a possibly insecure fabrication supply chain. While many techniques have focused on locking combinational logic (CL), an alternative latch-locking approach in which the sequential elements are locked has also gained significant attention. 
Latch (LAT) locking duplicates a subset of the flip-flops (FF) of a design, retimes these FFs and replaces them with latches, and adds two types of decoy latches to obfuscate the netlist. It then adds control circuitry (CC) such that all latches must be correctly keyed for the circuit to function correctly.  
This paper presents a two-phase attack on latch-locked circuits that uses a novel combination of deep learning, Boolean analysis, and integer linear programming (ILP). The attack requires access to the reverse-engineered netlist but, unlike SAT attacks, is oracle-less, not needing access to the unlocked circuit or correct input/output pairs. We trained and evaluated the attack using the ISCAS'89 and ITC'99 benchmark circuits. The attack successfully identifies a key that is, on average, 96.9\% accurate and fully discloses the correct functionality in 8 of the tested 19 circuits and leads to low function corruptibility (less than 4\%) in 3 additional circuits.
The attack run-times are manageable.  
\end{abstract}


\section{Introduction}
\label{sec:intro}
Modern integrated circuit (IC) design and manufacturing often rely on a global supply chain in which security 
and intellectual property (IP) rights are a significant concern. Logic locking (LL) is a common approach to protect hardware IP from an untrusted foundry \cite{Kamali2020InterLock, shamsi2019ip, yasin2017provably, Shakya2020CASlock, Kamali2019Fulllock}. 
In general, these approaches incorporate additional logic that supports the introduction of secret keys. When the user applies an incorrect key value to the locked circuit, the circuit outputs will often be wrong, thus locking the correct functionality.

Several LL approaches have been developed for  combinational \cite{yasin2016sarlock, yasin2017provably, Roy2022HOLL, 9586159} and sequential circuits 
\cite{5247148, Azar2021DataFlowObfus, Li2022JANUSHD, Tehranipoor2022oclock} as well as for scan chains 
\cite{8105900, 8709792, 9136991}.
Most of the proposed attacks on these defenses are oracle-based; they assume the attacker has access to a functioning unlocked circuit's primary inputs and outputs. These include attacks based on satisfiability checkers (SAT) \cite{7140252}, sensitization \cite{yasin2015improving}, removal and bypass \cite{xu2017novel, 8013714}, functional analysis \cite{8715163}, GF(2) linear algebra \cite{chen2021gfflush},
and various forms of machine learning (ML) \cite{8741028, 8942134, azar2020nngsat}.

In contrast, several oracle-less attacks have been proposed that require access to the netlist but not a functional circuit, making them, in some ways, more dangerous. These attacks take advantage of structural signatures associated with specific defenses \cite{sisejkovic2021logic,9006720, RANE2021Roshanisefat},
often using machine learning, such as OMLA~\cite{Alrahis2022OMLA}, GNNUnlock~\cite{9474039} and Snapshot~\cite{sisejkovic2021challenging}, etc~\cite{UNTANGLE2021Alrahis, Shamsi2022MLattackLUTLL, Swarup2021SCOPE, Alrahis2022GNNUnlockplus, Alrahis2022MuxLink}. Some attacks, like SAIL~\cite{8607163}, target structural signatures left by synthesis tools inserting XOR and other gate-level locking components. Other techniques focus on identifying the unique structural signatures of SAT-resilient logic \cite{9474039}. Many of these use ML classification as a critical step in identifying the secret key \cite{sisejkovic2021logic,9006720, sisejkovic2021challenging, 8607163}. Others, in contrast, are focusing on the identification of gates as either original or locking-related \cite{9474039,9530566,subhajitreverse}. None of these attacks, however, have been applied to locked latch-based circuits. 

Latch-Based Logic Locking (LBLL)~\cite{9300256}, referred to more simply as latch locking, is a less studied defense that aims to combine the merits of combinational and sequential logic locking. It first duplicates a subset of a design's FFs, then retimes them, and replaces them with latches. It then inserts two types of decoy latches, delay decoys and logic decoys, to obfuscate the netlist. 
After that, control circuitry is added such that all latches must be correctly keyed for the circuit to function correctly as a primary-secondary-based design. 
In particular, when correctly keyed, the delay decoys are forced to be transparent, and the logic decoys are forced to emit a constant 0. Additional combinational logic is added to ensure the 0 does not 
alter the correct operation of the circuit.
The approach demonstrates resilience to standard SAT and model-checking-based attacks \cite{9300256} and, to the best of our knowledge, has yet to be broken.  

This paper proposes an oracle-less attack on LBLL that combines deep learning, Boolean analysis, and integer linear programming. Our attack is based on the observation that the sequential graph associated with primary-secondary\footnote{In~\cite{9300256}, the authors adopt the terminology master and slave latches. We choose to use the terminology primary and secondary latches.} 
latch-based designs have a regular structure that is broken by the random insertion of decoy latches. This distinction yields structural signatures that can be taken advantage of by machine learning. The first phase of our attack identifies the logic decoys.
The logic decoys are then removed, and 
the circuit is simplified via constant propagation. In the second phase, the simplified circuit is input to a second ML classifier that identifies delay decoy latches. The softmax outputs of the classifier are fed as the coefficients in the objective function of an ILP whose constraints understand the correct structure of a primary-secondary latch-based design. The ILP finds a large pool of potential keys that are close to the classified output but also adhere to the latch constraints. We assume each of the identified keys can be independently evaluated by the attacker.

Our attack was trained and evaluated using the ISCAS’89 and ITC’99 benchmarks and configured to find a pool of 10k potential keys for each test circuit. The best key within the pool is on average 96.9\% accurate and measured the impact of incorrect keys by measuring its functional corruptibility \cite{9702267}. We found that the best-identified key unlocks the correct functionality in 8 of the tested 19 circuits and leads to low function corruptibility (less than 4\%) in 3 additional circuits.
The attack run times demonstrate the scalability of the approach, remaining less than 15 minutes in all circuits tested. 
%
%
%

A plethora of research tries to combine machine learning and constrained optimization in various ways, including using machine learning to speed up constrained optimization algorithms and end-to-end methods that feed the machine learning results into optimization algorithms~\cite{kotary2021endtoend}. In particular, this specific combination of classification with ILPs has been applied in natural language processing to detect disfluencies in speech~\cite{georgila2009using}. To the best of our knowledge, however, this is the first hardware security attack that combines the benefits of a trained ML classifier with an ILP. 

The remainder of this paper is organized as follows. Section~\ref{sec:backg} reviews latch locking and related background in machine learning. Section~\ref{sec:twophasedAttack} describes the proposed attack, including the motivation for our two-phase approach. Section~\ref{sec:expr} details experimental results together with two ablation studies that quantify the benefits of aspects of our attack. Some conclusions and opportunities for future work are discussed in the last section.

\section{Background}
\label{sec:backg}
\subsection{Latch-Based Logic Locking (LBLL)}
\begin{figure}[bt!]
    \centering  
    \includegraphics[width=\columnwidth]{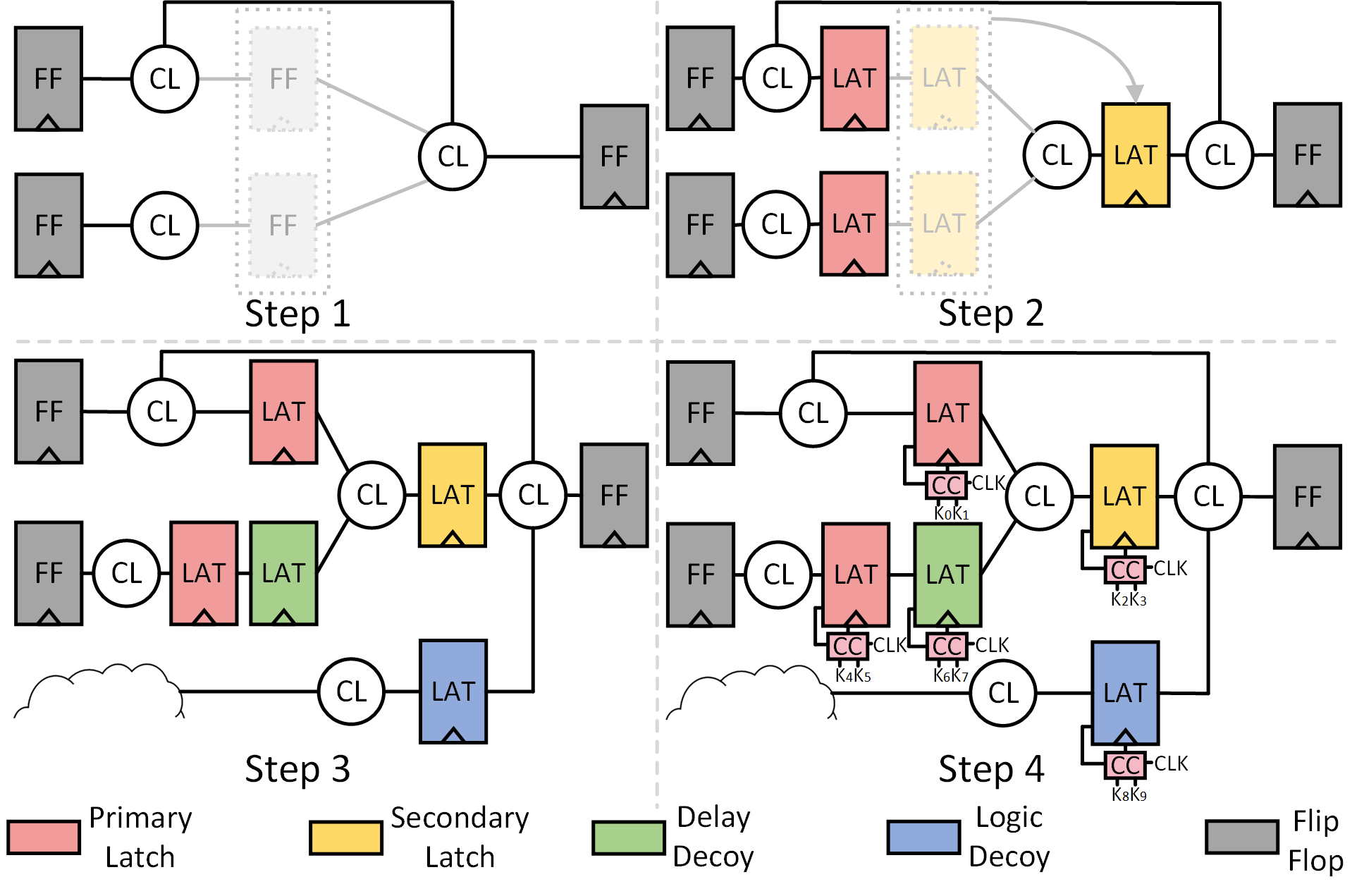}
    \caption{The four steps of latch-based logic locking. 
    Step 1: Select some interdependent FFs. Step 2: Duplicate FFs, retime some FFs and replace all duplicated FFs with latches. Step 3: Add delay decoys and logic decoys. Step 4: Add control circuitry.}
    \label{fig:LBLL_insert}
\end{figure}

Figure~\ref{fig:LBLL_insert} illustrates the four steps of latch locking. Step one uses a community detection algorithm to select a subset of FFs. Each selected FF is then duplicated and some of them are retimed (step 2) before being replaced with latches (step 3).\footnote{In principle, it is also possible to convert the FFs to latches before retiming, but commercial tools have better support for FF-based retiming.} 
Note, at this point, the latches must be two-colorable as alternating primary and secondary latches. Next, the latch locking script randomly inserts delay and logic decoy latches into the netlist to obfuscate the netlist and connects all latches to control circuitry (step 4). The control circuitry accepts two-bit keys to configure the four types of latches. In particular, delay decoys, when keyed correctly, are forced to be transparent and thus only influence the delay of the circuit. Logic decoys, on the other hand, when keyed correctly, output a fixed 0 value. The insertion of logic decoys must be coupled with extra OR/XOR/MUX gates to ensure the latch, when keyed correctly, does not corrupt the circuit's functionality. The possible cyclic connections and difficulty in setting the initial state complicate a SAT attack \cite{9300256}.

\begin{figure*}[h]
    \centering
    \includegraphics[width=0.8\textwidth]{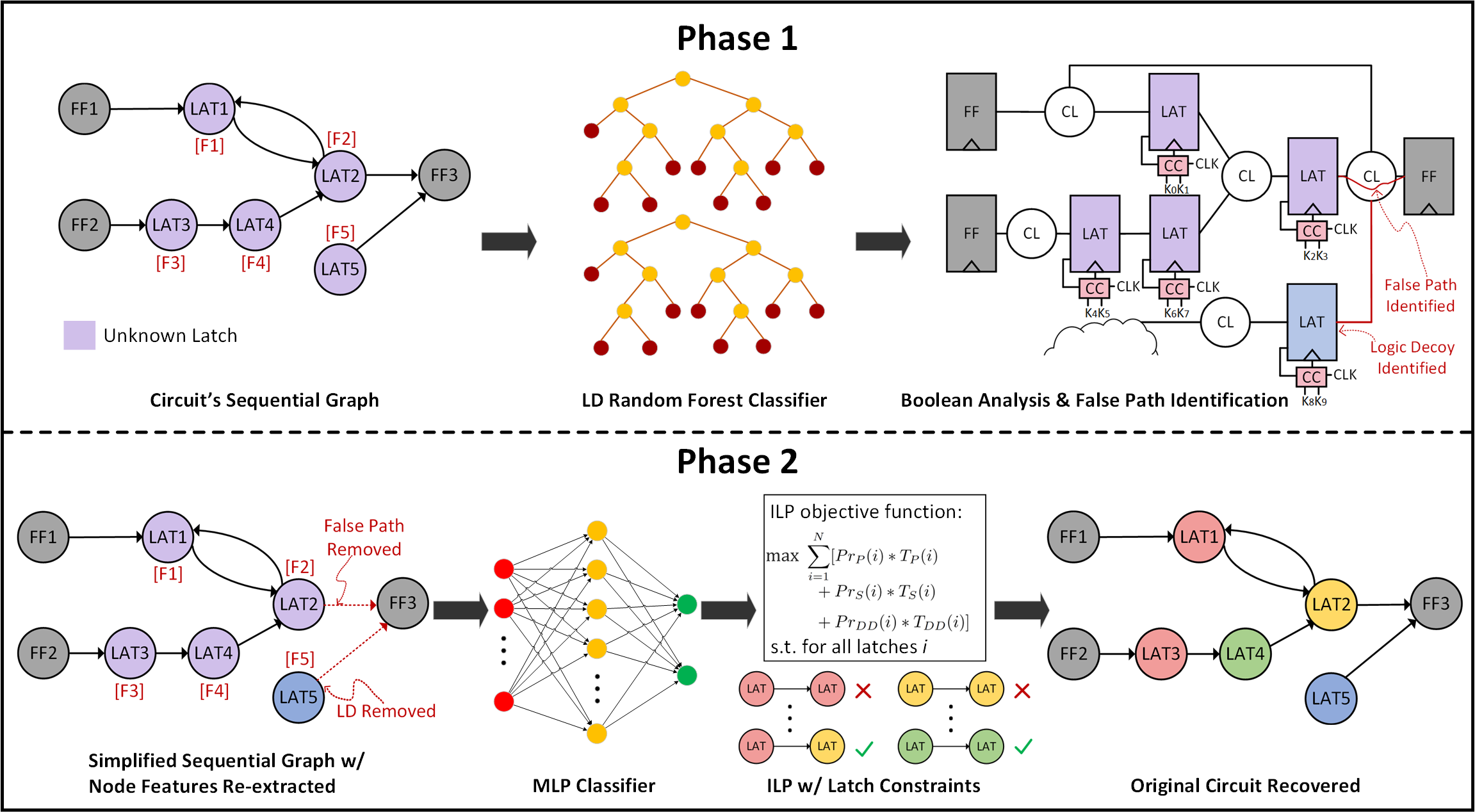}
    \caption{Overview of the proposed two-phase attack on LBLL.}
    \label{fig:attack_overview}
\end{figure*}

\subsection{Machine Learning Models}

\subsubsection{Multi-layer Perceptrons (MLP)}

A Multi-layer Perceptron (MLP) consists of an input, several hidden, and an output layer of fully connected neurons. The universal approximation theorem 
has proven that MLPs can learn any input-output function, motivating their widespread use in classification tasks. The softmax activation function is often used at the output layer to produce a probability for each class. 
Typically, the class with the highest probability is selected as the classification result.

\subsubsection{Random Forest (RF)}

A random forest (RF) is another widely used classifier that consists of many decision trees acting as an ensemble. Each decision tree recursively splits input samples based on features that lead to the smallest conditional entropy and output a classification vote. The class with the most votes from the ensemble of decision trees is selected as the final class prediction.

\subsection{Attacker Model}

Similar to most oracle-less attacks~\cite{9474039,8715163, 9006720}, we assume the adversary has access to the GDSII mask and has reverse-engineered the gate-level netlist.
The adversary also has information about the technology library and thus can estimate the circuit's static delays. Finally, we assume the attacker has detailed knowledge of the LBLL algorithm. We do not assume that the adversary has access to the unlocked circuit. That said, to evaluate the accuracy of our attack, we use the ground truth latch 
types and netlist.

\section{Proposed Two-Phase Attack}
\label{sec:twophasedAttack}

This section first provides motivation and an overview of our two-phase approach, as illustrated in Figure~\ref{fig:attack_overview}. It then formalizes our notion of a sequential graph, describes the 
features used in our ML classifiers and provides details of each attack phase.

\subsection{Motivation and Overview}
\label{sec:motiv}

\begin{figure}[H]
    \centering  
    \includegraphics[width=\columnwidth]{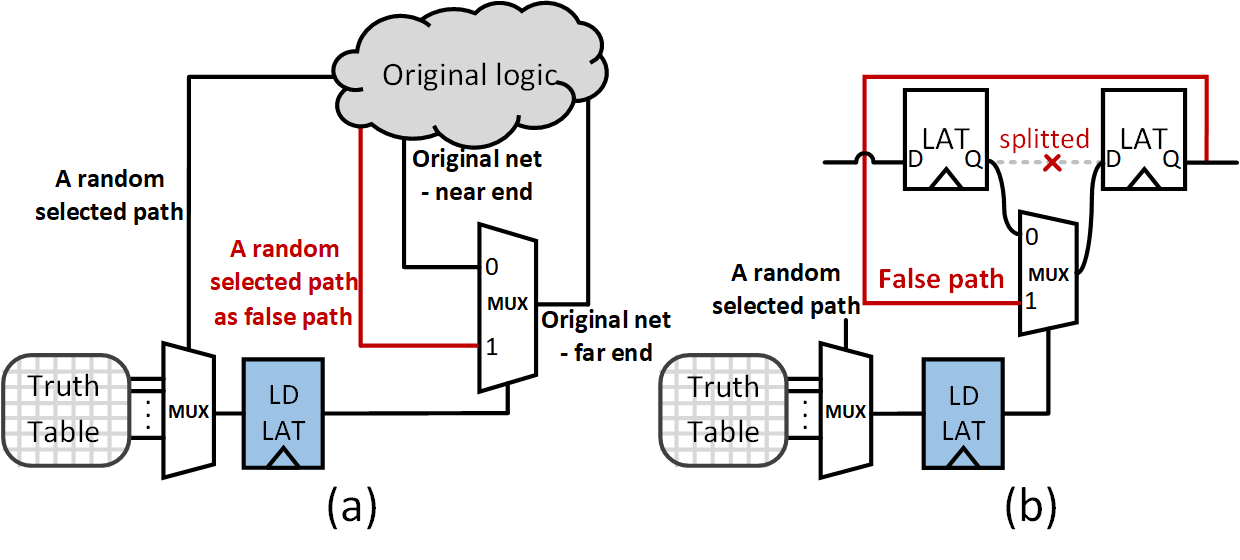}
    \caption{(a) Introduction of false paths by logic decoys. (b) A false path example.}
    \label{fig:false_path}
\end{figure}

The sequential graph of a circuit abstracts away the combinational logic of a circuit. Its nodes are primary input/outputs as well as sequential elements, flip-flops, and latches, and the edges represent the presence of combinational logic between nodes. 
The sequential graph of primary-secondary latch-based designs has a very regular structure; the graph is two-colorable. 
Our experiments have shown that it is relatively easy to identify the insertion of 
delay decoys that break this structure. However, 
the insertion of logic decoys
makes the circuit structure more complex. 
In particular, 
as shown in Figure~\ref{fig:false_path}, the 
insertion of logic decoys is sometimes coupled with the insertion of MUXes that create false paths between latches. 
In particular, the LBLL flow randomly inserts a MUX whose selection port is the output of a logic decoy. 
The “0” input to the MUX is attached to the near end of a randomly selected net that is cut.  
The output of the MUX is connected to the far end of the cut net. 
Therefore, when the logic decoy is keyed correctly, the MUX effectively re-connects the cut net and 
the circuit operates as if no decoy is added.
On the other hand, the “1” input to the MUX is randomly connected to another pin 
in the community, which can create a false path between latches, 
making the sequential graph's structure more complex.

This observation motivates our two-phase approach.
In our first phase (see Figure \ref{fig:attack_overview} top), we use ML to detect the logic decoys and then use Boolean analysis to remove the false paths created by these decoys. With the logic decoys and associated false paths removed, our second phase (see Figure \ref{fig:attack_overview} bottom) classifies the remaining latches using a combination of ML and ILP. In particular, we use the softmax outputs of a second classifier to create the objective function of an ILP whose constraints limit the solution space to legal primary-secondary configurations with delay decoys. The ILP solver is configured to not only find the closest legal key to that identified by the ML classifier but also identify many keys that are close to optimal, which the attacker can individually test.


\subsection{Sequential Graph and Node Feature Set}
\label{sec:features}

\begin{figure}[h]
    \centering  
    \includegraphics[width=\columnwidth]{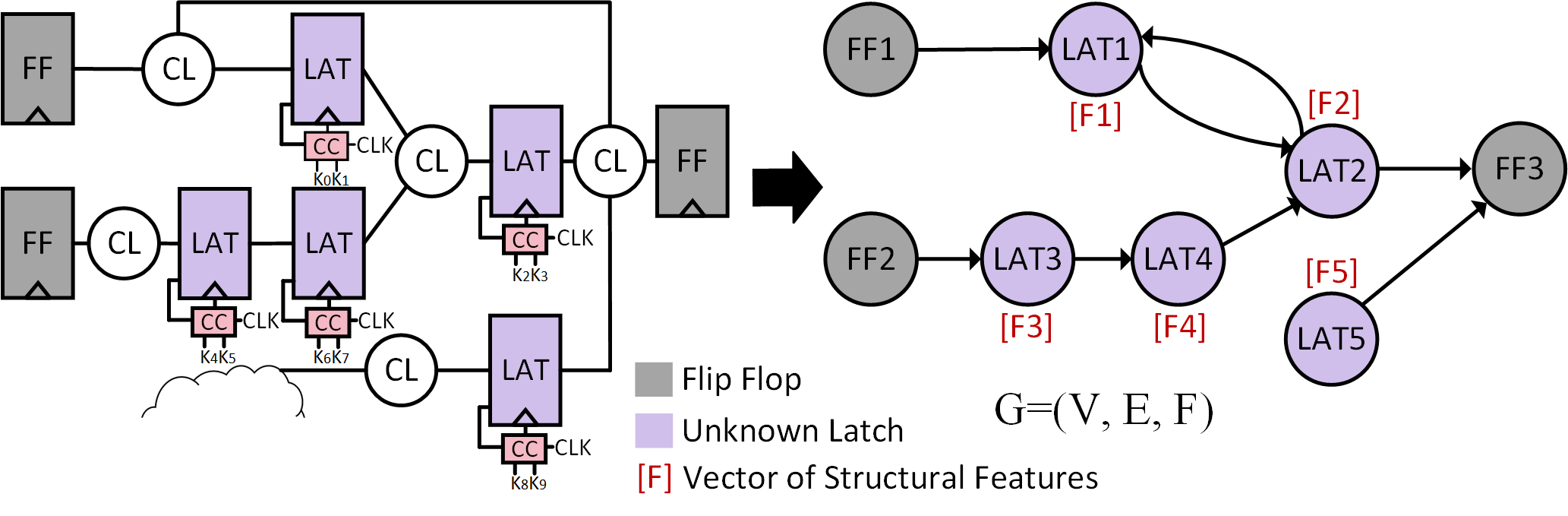}
    \caption{Illustration of generating a circuit's sequential graph.}
    \label{fig:graph_conversion}
\end{figure}

An example is the abstraction of a sequential graph from a circuit is illustrated in Figure~\ref{fig:graph_conversion}. 
A sequential graph is formally defined as $G = (V, E, F)$ where the set of nodes $V$ consists of latches, FFs, and primary inputs and outputs. An edge $e \in E$ exists between nodes if there is a combinational path between the node elements. $F$ represents a vector of structural features associated with each latch node used by our machine learning models to classify each latch individually. 

In the proposed approach, we extract fourteen features for each node latch, some of which are illustrated in Figure~\ref{fig:features}: 
\begin{figure}[h]
    \centering  
    \includegraphics[width=\columnwidth]{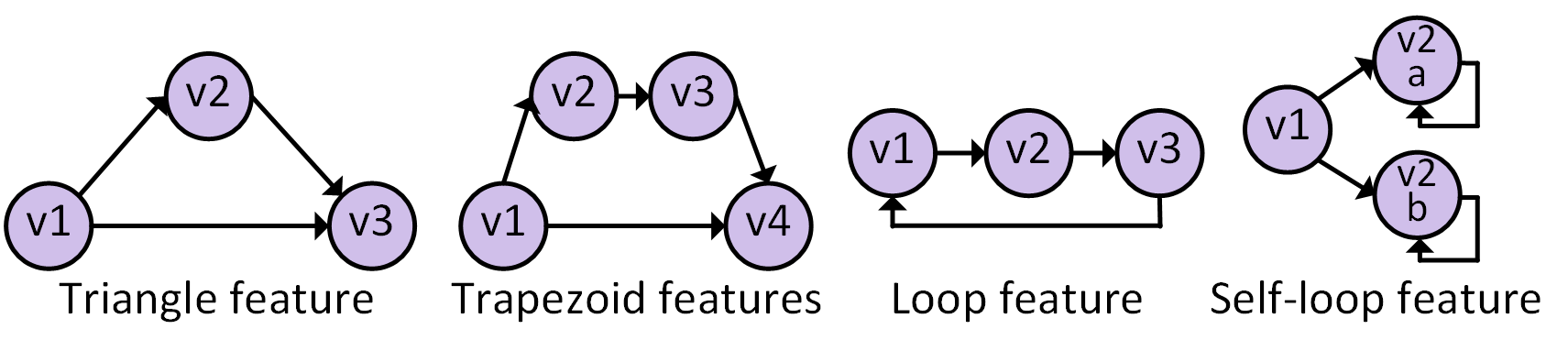}
    \caption{Several structural features identified.}
    \label{fig:features}
\end{figure}
\begin{enumerate}
  \item [1.] Triangle feature: the fraction of fan-ins $v_1$ of a node of interest (NOI) $v_2$ that share a fan-out.
  As defined below, it may detect decoys because this structure is not 2-colorable. 
  \begin{equation}
    \frac{\sum_{v_1 \in FI(v_2)} I[FO(v_2)\cap FO(v_1)]}{|FI(v_2)|}
\end{equation}
where $I$ denotes the indicator function, $FI$ and $FO$ return the fan-ins and fan-outs of a node.
  \item [2-3.] As an extension of the triangle feature, we define two trapezoidal features to detect when two consecutive decoy latches are inserted between primary and secondary latches. The first feature is 
    \begin{equation}
    \frac{\sum_{v_1 \in FI(v_2)} I[FO(FO(v_2))\cap FO(v_1)]}{|FI(v_2)|}
\end{equation}
where $v_2$ is the NOI. The second feature is similar but focuses on the fan-ins of fan-ins of the NOI (labeled $v_3$ in Figure \ref{fig:features}).
  \item [4-5.] Max fan-out delay and max fan-in delay, normalized for each circuit.
  \item [6.] Loop: a binary feature that detects if the NOI resides in a loop of three nodes, as shown in Figure~\ref{fig:features}
  \item [7.] Single fan-in or fan-out: a binary feature that detects if the NOI has only one fan-in or fan-out
  \item [8-10.] Three fan-in features: number of fan-in latches, FFs, and primary inputs.
  \item [11-13.] Three fan-out features: number of fan-out latches, FFs, and primary outputs.
  \item [14.] False self-loop feature: used for detecting logic decoy latches that introduce false self-loops in their fan-out latches, as shown in Figure~\ref{fig:features}, and defined as 
\begin{equation}
\max_{\substack{v_2 \in FO(v_1) \\ v_2 \in SL}}
    \frac{1}{|FI(v_2)|}
\end{equation}
where $SL$ is the set of latches that have a self-loop, and the max operation effectively yields the highest likelihood that the NOI $v_1$ is the cause of one of its fan-out $v_2$ to have a false self-loop.
\end{enumerate}
This set combines structural features that are specific to latch-based circuits with generic features that have been used in previous 
attacks~\cite{Alrahis2022OMLA, sisejkovic2021challenging, 9474039}. 

\subsection{Phase 1: Identify Logic Decoys}

Once the sequential graphs and feature vectors are extracted, phase 1 aims to identify the logic decoy latches and remove them. The circuit is then simplified via constant propagation. For the classifier in this phase, we found that a random forest outperforms other classifiers, including a support vector machine (SVM), MLP, and convolutional neural network (CNN), many of whom, for this problem, suffer from overfitting.

Note that this phase has a similar goal as the SAAM attack in \cite{sisejkovic2021deceptive} in that it, to some degree, is trying to identify randomly inserted MUXes. However, our approach is different because it focuses on the impact of insertion on the circuit's sequential graph and does not directly rely on the probabilities of the MUX input being connected. This means that our approach might still be effective even if LL was improved to incorporate their intelligent MUX insertion algorithm \cite{sisejkovic2021deceptive}. That said, guiding the LBLL MUX insertion step such that it can fool ML is an interesting area of future work.

\subsection{Phase 2: Identify Remaining Latches}
To train the second phase of our attack, we use ground-truth labels to remove logic decoys from the locked circuit, generate simplified sequential graphs and the associated feature vectors, then train a second classifier. For this classifier, we explore options with two and three output classes. The first distinguishes the delay decoys from primary/secondary latches, and the latter classifies all three types.
In this phase, we use an MLP whose output activation function is a softmax to yield probabilities for each class. These probabilities are used as the coefficients in the ILP objective function and guide the optimization process. 

For the ILP, two sets of binary variables, $T$ and $C$, are used. Each latch is associated with three $T$ variables, $T_P$, $T_S$, and $T_{DD}$, as logic decoys are presumably already identified and removed in phase 1. A $T$ variable equal to 1 indicates that the latch belongs to the corresponding type ($T_P(i)=1$ indicates that the $i^{th}$ latch is classified as a primary latch). Each latch also has one $C$ variable, which specifies its color, whose related constraints will be explained later in this section. The ILP objective function is to maximize 
  \begin{equation*}
    \sum_{i=1}^{N} [Pr_P(i)\cdot T_P(i)+Pr_S(i)\cdot T_S(i)+Pr_{DD}(i)\cdot T_{DD}(i)]
\end{equation*}
Where $N$ is the number of latches and $Pr_P$, $Pr_S$, and $Pr_{DD}$ are the softmax probabilities from the MLP classifier. Note that $Pr_P=Pr_S=Pr_{PS}$ in our two-level classifier.

We generate three sets of constraints for the ILP, as described below.

\subsubsection{Fundamental Constraints}

We refer to the first set of constraints as fundamental, as follows 

\begin{enumerate}[start=0,label={\bfseries F\arabic*:}]
  \item  $T_P (i)+T_S (i)+T_{DD} (i)=1$
  \item If $T_P (i)=1$, then $C(i)=1$
  \item If $T_S (i)=1$, then $C(i)=0$
\end{enumerate}

The \textbf{F0} constraint ensures that each latch is classified into precisely one type of latch.
The next two constraints, \textbf{F1} and \textbf{F2}, correlate the latch's $T$ variable to its color $C$.
In particular, the $C$ variable of every primary latch is equal to 1, and that of every secondary latch is equal to 0. 

Note that there are no such constraints for DD latches, which means that the $C$ variable of a DD latch can either be 1 or 0. Because of the potential existence of delay decoy latches between primary and secondary latches, the color variable $C$ is used to capture the identity of the neighboring primary/secondary latches along the path, as described next. 

\begin{figure}[htbp]
    \centering  
    \includegraphics[width=0.6\columnwidth]{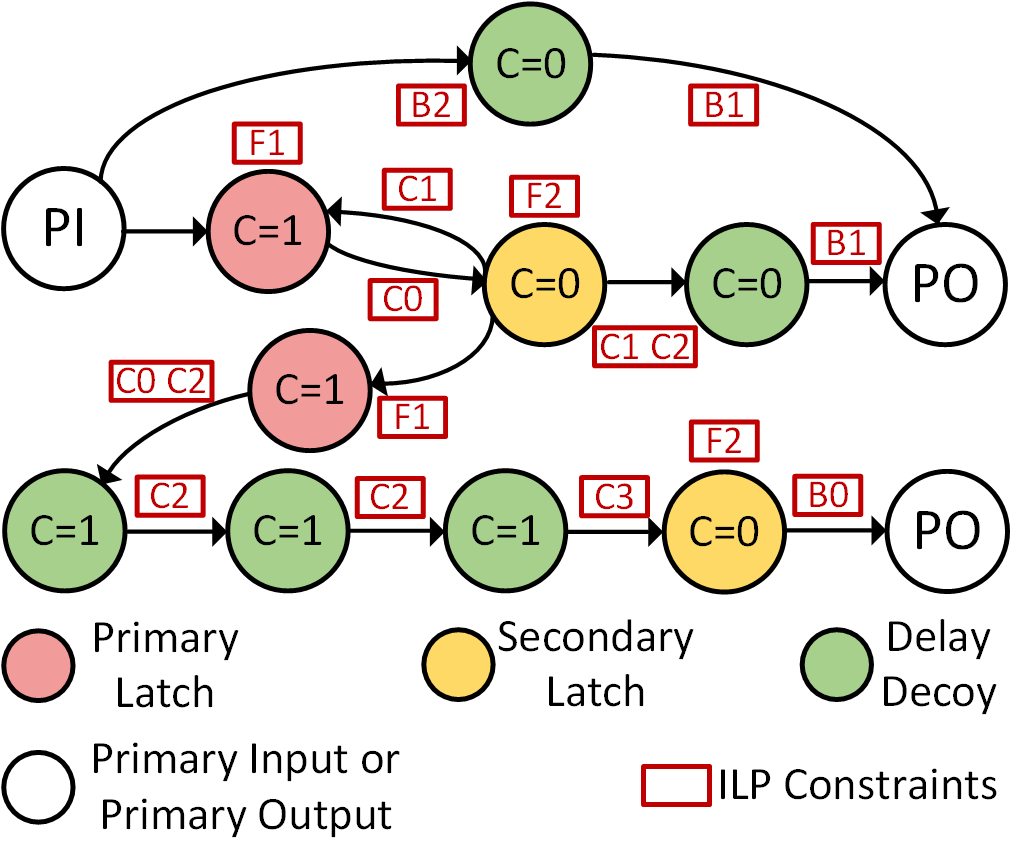}
    \caption{Illustration of ILP constraints.}
    \label{fig:constraint}
\end{figure}

\subsubsection{Coloring Constraints}

The second set of constraints is for coloring and is applied to every pair of neighbor latches, where latch $i$ drives latch $j$. This set of constraints aims to constrain solutions to those in which the primary and secondary latches are placed alternately.

\begin{enumerate}[start=0,label={\bfseries C\arabic*:}]
  \item If $T_P(i)=1$, then $T_P(j)=0$
  \item If $T_S(i)=1$, then $T_S(j)=0$
 \end{enumerate}
%

Recall that primary and secondary latches are created by splitting the FFs, the neighbor latches of a primary latch must be secondary latches or delay decoys, and the neighbor latches of a secondary latch must be primary latches or delay decoys. In other words, a primary latch cannot follow a primary latch. Similarly, a secondary latch cannot be followed by a secondary latch.
%
%

Moreover, 
with only the two constraints above that ensure primary and secondary latches cannot be neighbors with the same type of latches, there still can be cases where two primary latches or two secondary latches are each other's non-decoy neighbors, which should also be avoided. To address this issue, we propose two more coloring constraints. 

\begin{enumerate}[start=2,label={\bfseries C\arabic*:}]
  \item If $T_{DD}(j)=1$, then $C(j)=C(i)$
  \item If $T_{DD}(i)=1$ and $T_P(j)+T_S(j)=1$, then $C(j)\neq C(i)$
 \end{enumerate}
 
The \textbf{C2} constraint ensures the color of delay decoy latches will be the same as their driving latch, effectively passing the color information to any fan-out latch. The \textbf{C3} constraint guarantees that the color of a primary/secondary fan-out latch will be different from that of a driving delay decoy. Together, these two constraints guarantee that the color of a primary/secondary latch after a delay decoy chain will differ from the color of the primary/secondary latch before the delay decoy chain. 

An example of the impact of these constraints is illustrated in Figure~\ref{fig:constraint}. It shows a primary latch 
(with C=1) driving a chain of delay decoys followed by a secondary latch (with C=0). 
The color of the delay decoy that is the immediate fan-out of the primary latch is 1 because the delay decoy carries the color of the previous non-decoy latch (due to \textbf{C2}). The second delay decoy also has the same color as the first delay decoy to pass the information that the previous non-decoy latch is a primary (also due to \textbf{C2}). In other words, the delay decoys here act as ``pseudo primary latches'' as they have the same color as the primary latches. The last pair of neighbor latches is a delay decoy driving a secondary latch. The delay decoy, which can be regarded as a ``pseudo primary latch'', indicates that the next non-decoy latch should be a secondary instead of a primary latch (due to constraint \textbf{C3}). 

\subsubsection{Boundary Constraints}

The third set of constraints is the latch boundary constraints which helps disambiguate the coloring options and avoids the assignment of primary and secondary to be switched. Since the community-based algorithm applied in LBLL backtracks from the largest fan-in cone \cite{9300256}, the sub-graph near the POs has less complexity than near the PIs. Thus, we select the primary output constraint to direct the coloring. 
\begin{enumerate}[start=0,label={\bfseries B\arabic*:}]
  \item If latch $i$ is connect to a PO and $T_{DD}(i)=0$, then $T_S(i)=1$
  \item If latch $i$ is connect to a PO and $T_{DD}(i)=1$, then $C(i)=0$
  \item If latch $i$ is connected to a PI and PO, then $T_{DD}(i)=1$
  \end{enumerate}

The first boundary constraint \textbf{B0} is motivated by the observation that every flip-flop is divided into a pair of primary and secondary latches. That is to say, there must be a secondary latch after every primary latch. 
Therefore, if a latch is immediately connected to PO and is not a delay decoy, it must be a secondary latch, as illustrated in Figure~\ref{fig:constraint}. Conversely, if a latch immediately connected to a PO is a delay decoy, then its $C$ variable must be 0 to ensure any primary/secondary latch that drives it is classified as a secondary (see \textbf{B1}). This is because the color of delay decoys is always consistent with that of its fan-in latches as enforced by the coloring constraints. Lastly, if a latch is connected to both a PI and a PO, it can neither be a primary nor secondary because they appear in pairs (see \textbf{B2}). Thus, such a latch must be a delay decoy with a color equal to 0.
Because FFs have similar boundary effects, we treat fan-in FFs to latches as PIs and fan-out FFs to latches as POs.


 \section{Experimental Results}
\label{sec:expr}
\subsection{Experiment Setup}
The proposed two-phase attack was evaluated on ISCAS'89 and ITC'99 benchmark circuits\footnote{We added reset to the ITC'99 benchmark circuits.}. The netlist to graph and sequential graph extraction was implemented in Python with the NetworkX library. The model training, inference, and subsequent ML analysis were implemented using Pytorch and scikit-learn. The Boolean analysis was performed in the Cadence Genus tool. All experiments, other than the ILP components, were performed on an Intel i7-8700 CPU running at 3.20 GHz with 16-GB RAM and NVIDIA GeForce RTX 2080 GPU with 16-GB memory. The ILPs were conducted on an Intel i7-10850H CPU running at 2.70 GHz with 32-GB RAM.
\subsection{Dataset Generation}
In total, we tested 19 circuits across the two benchmark suites. For each circuit, we locked it with LBLL scripts and generated 11 locked variants with different random seeds.\footnote{We thank the authors of \cite{9300256} for making their scripts available to us.} 
To attack each of the 19 benchmark circuits, we trained a model using
the 18 other circuits and their variants, splitting samples between training and validation. The number of latch samples for training the models ranged from 75k to 77k.
For the first classifier (phase 1), we trained the models with the original locked circuits.Then, we trained the models with simplified circuits for the second classifier (phase 2), removing the ground-truth logic decoys and associated false paths.
\subsection{Accuracy and Run-Time Results}
The first-phase models achieve an average accuracy of 98.1\%. For the second phase, 
Columns 3-4 in Table~\ref{tab: results_1} and~\ref{tab: results_2} details the overall 
accuracy under four different MLP and ILP configurations used in the second phase. In particular, we tested both a 2-level and 3-level classifier and, in both cases, configured the ILP to search for the top 1 (labeled ``T-1'') and top 10k (labeled ``T-10k'') potential classifications of latches. We report the accuracy of the only/best-identified classifications. 

For both top-1 and top-10k results, the latch constraints with the 2-level classifier yielded the highest average accuracy, fully disclosing the specified secret key in 6 of 19 circuits and, on average, achieving keys that are, on average, 96.9\% accurate.

Note that even for the largest circuit tested, the MLP inference run-time is less than 5 minutes and the ILP completes its search in less than 15 minutes.\footnote{To aid reproducibility, we have posted our attack code on the web page https://github.com/siriuscdk/LBLL}

\begin{table}[tbhp]
\footnotesize
\centering
\begin{tabular}{|l|l|l|l|l|l|}
\hline
\multirow{2}{*}{Circuit} &
\# of &
\multicolumn{3}{c|}{3-Level MLP  (\%)} \\
& keys  & T-1  & T-10k  & FC 
\\ \hline
s298              & 100   & 85.0  & 88.0 &  8.6   \\ \hline
s9234             & 188   & 87.2    & 94.1  & 0.02 \\ \hline
s13207            & 116  & 98.3    &100.0  &  0  \\ \hline
s15850            & 252    & 89.7   & 96.0 &  0  \\ \hline
s35932            & 592    & 90.5    & 91.6 & 3.2  \\ \hline
s38417            & 1060  & 97.5   & 99.0 &  17.4  \\ \hline
s38584            & 452   & 93.8  &97.8 &  8.5 \\ \hline
b03               & 124   & 83.1   & 93.5 &  19.8 \\ \hline
b04               & 112   & 100.0  & 100.0 &  0   \\ \hline
b07               & 412  & 90.8  & 95.1 & 15.1  \\ \hline
b11               & 412  & 91.3   & 96.1    &  34.1 \\ \hline
b12               & 424  & 94.6  & 97.9 & 0 \\ \hline
b13               & 156   & 97.4    & 100.0 &  0 \\ \hline
b14               & 984   & 97.5  & 98.3 & 3.3 \\ \hline
b15               & 2736 & 94.6  & 95.0 & 6.6 \\ \hline
b17               & 3858  & 92.0   & 92.5 &  7.9 \\ \hline
b20               & 940   & 99.8   & 100.0  &  0 \\ \hline
b21               & 840   & 99.0    & 100.0   &  0  \\ \hline
b22               & 920   & 99.6   & 100.0  &  0  \\ \hline
Ave. &      & 93.8 & 96.6 & 6.6  \\ \hline
\end{tabular}

\caption{Attack accuracy results for 3-Level MLP.}
\label{tab: results_1}
\end{table}

\begin{table}[H]
\footnotesize
\centering
\begin{tabular}{|l|l|l|l|l|l|}
\hline
\multirow{2}{*}{Circuit} &
\# of &
\multicolumn{3}{c|}{2-Level MLP  (\%)} \\

& keys  & T-1  & T-10k   & FC 
\\ \hline
s298              & 100  & 84.0   &  88.0  & 8.6    \\ \hline
s9234             & 188  & 87.2   &  94.1  & 0.03 \\ \hline
s13207            & 116  & 98.3   &  100.0 & 0.0 \\ \hline
s15850            & 252  & 91.7   &  97.2  &  0.0 \\ \hline
s35932            & 592  & 90.7   & 92.7  & 1.5   \\ \hline
s38417            & 1060 & 97.3   &  99.1  & 17.3 \\ \hline
s38584            & 452  & 93.8   & 97.6   &  8.5 \\ \hline
b03               & 124  & 89.5   & 93.5   &  19.8  \\ \hline
b04               & 112  & 100.0  & 100.0  &  0.0    \\ \hline
b07               & 412  & 91.7   & 95.6   & 10.6   \\ \hline
b11               & 412  & 93.0   & 96.8   & 46.8 \\ \hline
b12               & 424  & 96.5   & 99.3   & 0.0 \\ \hline
b13               & 156  & 96.2   & 100.0  &  0.0  \\ \hline
b14               & 984  & 96.8   & 98.1   & 3.3  \\ \hline
b15               & 2736 & 95.4   & 95.9   & 6.6 \\ \hline
b17               & 3858 & 92.9   & 93.5   & 8.2 \\ \hline
b20               & 940  & 99.1   & 100.0  & 0.0 \\ \hline
b21               & 840  & 99.3   & 100.0  & 0.0    \\ \hline
b22               & 920  & 99.6   & 100.0  & 0.0  \\ \hline
Ave. &      & 94.4 & 96.9 & 6.9  \\ \hline
\end{tabular}

\caption{Attack accuracy results for 2-Level MLP.}
\label{tab: results_2}
\end{table}


\subsection{Functional Corruptibility Analysis}

We also measured the functional corruptibility of each circuit with the best-identified key in the two configurations whose accuracy is shown in Table~\ref{tab: results_1} and~\ref{tab: results_2}. The functional corruptibility of a keyed combinational circuit is the fraction of output bits that are incorrect~\cite{6616532, mertensat}. 
For sequential circuits, however, the inputs should be randomly selected and sequentially applied over $b$ clock cycles to enable errors in the next state logic to propagate to the primary outputs \cite{9702267}. We chose $b=1000$ with different random inputs 1,000 times and averaged the results to obtain a more comprehensive measure of average functional corruptibility.

The results are shown in the FC columns in Table~\ref{tab: results_1} and~\ref{tab: results_2}. For several cases, even a small number of incorrect key bits lead to a large fraction of output errors, as may be expected for a sequential circuit. However, overall, 8 circuits achieve 100\% correct functionality (FC=0). An FC=0 is  expected for keys that are 100\% accurate.
Interestingly, however, in some cases, FC=0 for keys that are less than 100\% accurate.
We manually investigated these cases and found that at least some misclassified latches are actually functionally redundant because all their fanouts are logic decoys. We believe this is a consequence of the locking script sequentially adding decoys without verifying that the decoy is not functionally redundant. 
%
%
For many other cases, the relatively low average functional corruptibility indicates that the identified keys, even if not perfectly accurate, disclose most of the circuit's functionality.

\subsection{Feature Importance Analysis}
\begin{figure}[htbp]
    \centering  
    \includegraphics[width=\columnwidth]{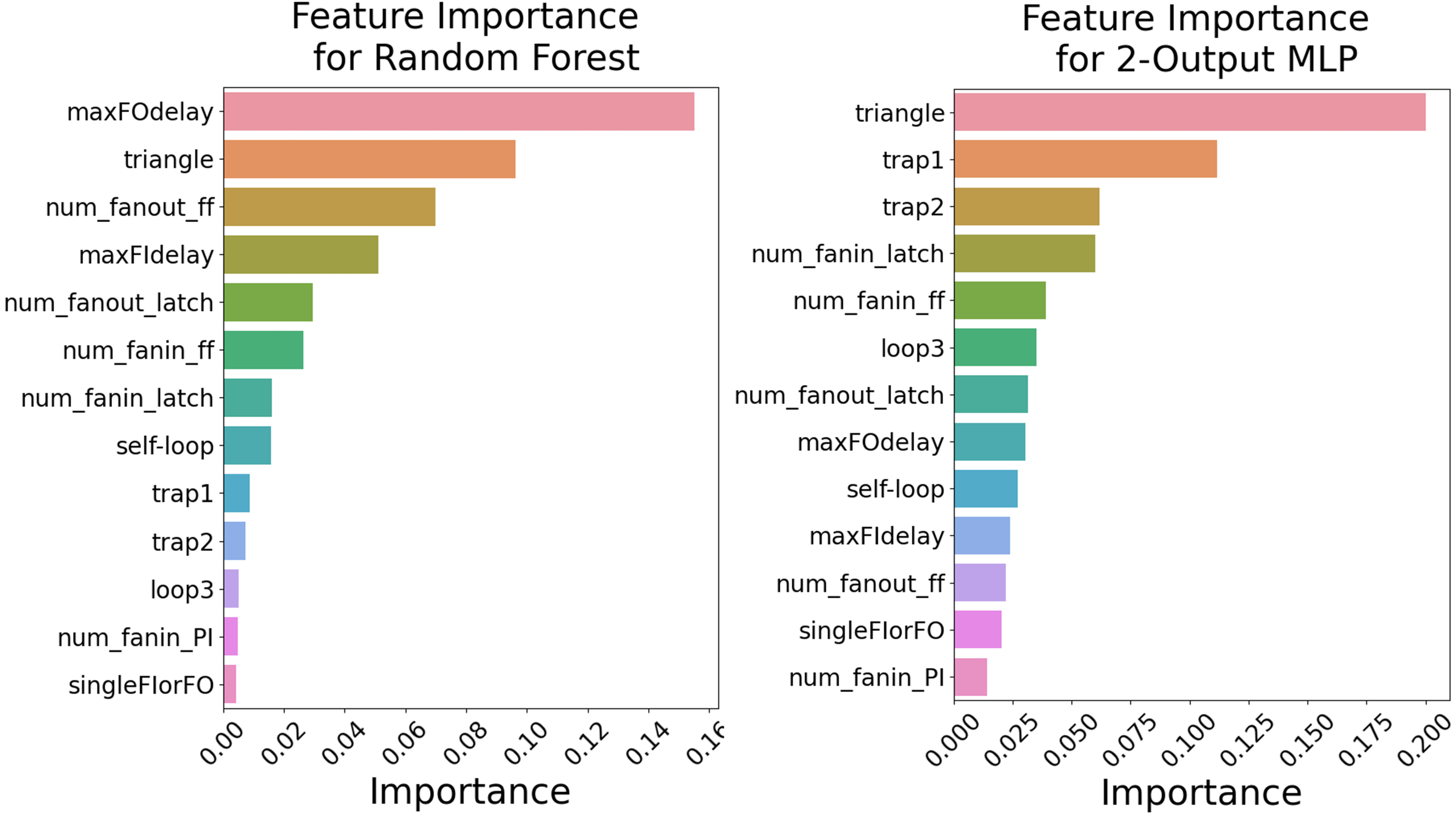}
    \caption{Feature importance for the two ML classifiers.}
    \label{fig:feature_importance}
\end{figure}
We performed feature importance analysis for the two classifiers used in our attack, as shown in Figure~\ref{fig:feature_importance}.
For the RF classifier, which is used to detect logic decoy latches, the max fan-out delay, triangle, and the number of fan-out FFs are the most salient features. Logic decoys usually have relatively small fan-out delay compared to other latches, and most of them have zero fan-out FFs.
The triangle and two trapezoid features are the most important for the second MLP classifier and are used to detect delay decoys. This result makes sense because these three features help in detecting insertions of individual and pairs of decoy latches that create non-2-colorable sub-graphs.

\subsection{Baseline MLP and Ablation Studies}

To further justify our approach, we first created a baseline, one-stage 4-class MLP to classify all latches with the features described in Section \ref{sec:features} and obtained an average accuracy of 82.99\%. We conducted two ablation studies to quantify the value of different components of our attack, as shown below.

\subsubsection{Value of ILP}

To quantify the value of our ILP, we replaced it with a recursive search algorithm that found the 10k \emph{closest} keys to the MLP identified result, where we used the MLP softmax probabilities to define the (weighted) distance between keys. The average accuracy for this algorithm is 87.4\%. The resulting drop in accuracy of around 9\% 
illustrates the significant benefit of the ILP limiting the search space to properly colorable classifications.

\subsubsection{Number of Classifiers}

To quantify the advantage of the two-phase approach over a single-phased approach, we combined the baseline 4-level MLP with false path removal and an ILP. After the classifier, we removed the identified logic decoys and simplified the circuit. We then formulated a T-10k ILP with the simplified circuit and the objective function coefficients from the MLP. The average accuracy for this approach is 89.8\%. The result 
shows an overall degradation in an average accuracy of around 7\%.

\section{Conclusions and future work}

This paper presents an oracle-less attack on latch-based logic locking (LBLL). Our empirical results show that the best-identified keys are, on average, 96.9\% accurate and the correct functionality is fully or mostly disclosed in the majority of the circuits tested. The attack run times are all less than 15 minutes.
%
%
Even though the secret key and associated functionality sometimes remain partially hidden, our results show that the current structure of LBLL circuits can leak important information to an attacker. This attack motivates improvements in latch locking to better obfuscate the structure of the locked circuits. These improvements can include (i) actively avoiding the information leakage of structures closely adhering to the identified coloring constraints and (ii) introducing other types of logic decoys, such as logic decoys that emit a logic one instead of a logic zero.
 
Since the proposed attack is based on graph analysis, we also note that graph-based methods may complement our node-based approach. This may include characterizing and classifying paths with two or three latches. Moreover, while our classifiers rely on specific structural signatures that we manually identified, identifying these features automatically using other forms of ML is interesting future work. In addition, another interesting direction to explore is to 
add our novel constraints to more traditional 
oracle-aided sequential SAT (Boolean Satisfiability) attacks 
of unrolled LBLL circuits.
%
Finally, we note that our attack illustrates the benefits of combining data-driven and structural analyses and assert that the proposed combination of MLP and ILP is a good template for many CAD problems.

\bibliographystyle{IEEEtran}
\bibliography{reference}

\begin{thebibliography}{10}
\providecommand{\url}[1]{#1}
\csname url@samestyle\endcsname
\providecommand{\newblock}{\relax}
\providecommand{\bibinfo}[2]{#2}
\providecommand{\BIBentrySTDinterwordspacing}{\spaceskip=0pt\relax}
\providecommand{\BIBentryALTinterwordstretchfactor}{4}
\providecommand{\BIBentryALTinterwordspacing}{\spaceskip=\fontdimen2\font plus
\BIBentryALTinterwordstretchfactor\fontdimen3\font minus
  \fontdimen4\font\relax}
\providecommand{\BIBforeignlanguage}[2]{{%
\expandafter\ifx\csname l@#1\endcsname\relax
\typeout{** WARNING: IEEEtran.bst: No hyphenation pattern has been}%
\typeout{** loaded for the language `#1'. Using the pattern for}%
\typeout{** the default language instead.}%
\else
\language=\csname l@#1\endcsname
\fi
#2}}
\providecommand{\BIBdecl}{\relax}
\BIBdecl

\bibitem{Kamali2020InterLock}
H.~M. Kamali, K.~Z. Azar, H.~Homayoun, and A.~Sasan, ``{InterLock: An
  Intercorrelated Logic and Routing Locking},'' in \emph{2020 ACM/IEEE
  International Conference On Computer Aided Design (ICCAD)}, 2020.

\bibitem{shamsi2019ip}
K.~Shamsi, M.~Li, K.~Plaks, S.~Fazzari, D.~Z. Pan, and Y.~Jin, ``{{IP}
  Protection and Supply Chain Security through Logic Obfuscation: A Systematic
  Overview},'' \emph{ACM Transactions on Design Automation of Electronic
  Systems (TODAES)}, vol.~24, no.~6, pp. 1--36, 2019.

\bibitem{yasin2017provably}
M.~Yasin, A.~Sengupta, M.~T. Nabeel, M.~Ashraf, J.~Rajendran, and O.~Sinanoglu,
  ``{Provably-Secure Logic Locking: From Theory To Practice},'' in
  \emph{Proceedings of the 2017 ACM SIGSAC Conference on Computer and
  Communications Security}, 2017, pp. 1601--1618.

\bibitem{Shakya2020CASlock}
B.~Shakya, X.~Xu, M.~Tehranipoor, and D.~Forte, ``{CAS-lock: A
  Security-Corruptibility Trade-off Resilient Logic Locking Scheme},'' in
  \emph{2020 IACR Trans. on Cryptographic Hardware and Embedded Systems
  (CHES)}, 2020, p. 175–202.

\bibitem{Kamali2019Fulllock}
H.~M. Kamali, K.~Z. Azar, H.~Homayoun, and A.~Sasan, ``{Full-Lock: Hard
  Distributions of SAT instances for Obfuscating Circuits using Fully
  Configurable Logic and Routing Blocks},'' in \emph{2019 ACM/IEEE Design
  Automation Conference (DAC)}, 2019.

\bibitem{yasin2016sarlock}
M.~Yasin, B.~Mazumdar, J.~J. Rajendran, and O.~Sinanoglu, ``{SARLock}: {SAT}
  attack resistant logic locking,'' in \emph{2016 IEEE International Symposium
  on Hardware Oriented Security and Trust (HOST)}, 2016.

\bibitem{Roy2022HOLL}
G.~Takhar, R.~Karri, C.~Pilato, and S.~Roy, ``{HOLL}: Program synthesis for
  higher order logic locking,'' in \emph{Tools and Algorithms for the
  Construction and Analysis of Systems}, D.~Fisman and G.~Rosu, Eds.\hskip 1em
  plus 0.5em minus 0.4em\relax Cham: Springer International Publishing, 2022,
  pp. 3--24.

\bibitem{9586159}
M.~R. Muttaki, R.~Mohammadivojdan, M.~Tehranipoor, and F.~Farahmandi,
  ``{HL}ock: Locking ips at the high-level language,'' in \emph{2021 58th
  ACM/IEEE Design Automation Conference (DAC)}, 2021, pp. 79--84.

\bibitem{5247148}
R.~S. Chakraborty and S.~Bhunia, ``{{HARPOON}: An Obfuscation-Based SoC Design
  Methodology for Hardware Protection},'' \emph{IEEE Transactions on
  Computer-Aided Design of Integrated Circuits and Systems}, 2009.

\bibitem{Azar2021DataFlowObfus}
K.~Z. Azar, H.~M. Kamali, S.~Roshanisefat, H.~Homayoun, C.~P. Sotiriou, and
  A.~Sasan, ``{Data Flow Obfuscation: A New Paradigm for Obfuscating
  Circuits},'' \emph{IEEE Transactions on Very Large Scale Integration (VLSI)
  Systems}, vol.~29, no.~4, pp. 643 -- 656, 2021.

\bibitem{Li2022JANUSHD}
L.~Li and A.~Orailoglu, ``{JANUS-HD: Exploiting FSM Sequentiality and Synthesis
  Flexibility in Logic Obfuscation to Thwart SAT Attack While Offering Strong
  Corruption},'' in \emph{2022 Design Automation \& Test in Europe Conf.
  (DATE)}, 2022.

\bibitem{Tehranipoor2022oclock}
\BIBentryALTinterwordspacing
M.~S. Rahman, R.~Guo, H.~M. Kamali, F.~Rahman, F.~Farahmandi, M.~Abdel-Moneum,
  and M.~Tehranipoor, ``O'clock: Lock the clock via clock-gating for soc ip
  protection,'' in \emph{Proceedings of the 59th ACM/IEEE Design Automation
  Conference}, ser. DAC '22.\hskip 1em plus 0.5em minus 0.4em\relax New York,
  NY, USA: Association for Computing Machinery, 2022, p. 775–780. [Online].
  Available: \url{https://doi.org/10.1145/3489517.3530542}
\BIBentrySTDinterwordspacing

\bibitem{8105900}
X.~Wang, D.~Zhang, M.~He, D.~Su, and M.~Tehranipoor, ``{Secure Scan and Test
  Using Obfuscation Throughout Supply Chain},'' \emph{IEEE Trans. on
  Computer-Aided Design of Integrated Circuits and Systems}, 2018.

\bibitem{8709792}
R.~Karmakar, S.~Chattopadhyay, and R.~Kapur, ``{A Scan Obfuscation Guided
  Design-for-Security Approach for Sequential Circuits},'' \emph{IEEE
  Transactions on Circuits and Systems II: Express Briefs}, vol.~67, no.~3, pp.
  546--550, 2020.

\bibitem{9136991}
S.~Potluri, A.~Aysu, and A.~Kumar, ``{SeqL: Secure Scan-Locking for IP
  Protection},'' in \emph{2020 21st International Symposium on Quality
  Electronic Design (ISQED)}, 2020, pp. 7--13.

\bibitem{7140252}
P.~{Subramanyan}, S.~{Ray}, and S.~{Malik}, ``Evaluating the security of logic
  encryption algorithms,'' in \emph{2015 IEEE International Symposium on
  Hardware Oriented Security and Trust (HOST)}, 2015.

\bibitem{yasin2015improving}
M.~Yasin, J.~J. Rajendran, O.~Sinanoglu, and R.~Karri, ``{On Improving the
  Security of Logic Locking},'' \emph{IEEE Transactions on Computer-Aided
  Design of Integrated Circuits and Systems}, 2016.

\bibitem{xu2017novel}
X.~Xu, B.~Shakya, M.~M. Tehranipoor, and D.~Forte, ``{Novel Bypass Attack and
  BDD-based Tradeoff Analysis Against all Known Logic Locking Attacks},'' in
  \emph{International Conference on Cryptographic Hardware and Embedded
  Systems}.\hskip 1em plus 0.5em minus 0.4em\relax Springer, 2017, pp.
  189--210.

\bibitem{8013714}
M.~Yasin, B.~Mazumdar, O.~Sinanoglu, and J.~Rajendran, ``{Removal Attacks on
  Logic Locking and Camouflaging Techniques},'' \emph{IEEE Transactions on
  Emerging Topics in Computing}, 2020.

\bibitem{8715163}
D.~Sirone and P.~Subramanyan, ``{Functional Analysis Attacks on Logic
  Locking},'' in \emph{2019 Design Automation and Test in Europe Conference
  Exhibition (DATE)}, 2019.

\bibitem{chen2021gfflush}
D.~Chen, C.~Lin, and P.~A. Beerel, ``{{GF}-Flush: A {GF}(2) Algebraic Attack on
  Secure Scan Chains},'' in \emph{2021 IEEE Int. Symp. on Defect and Fault
  Tolerance in VLSI and Nanotechnology Systems (DFT)}, 2021.

\bibitem{8741028}
P.~Chakraborty, J.~Cruz, and S.~Bhunia, ``{{SURF}: Joint Structural Functional
  Attack on Logic Locking},'' in \emph{2019 IEEE International Symposium on
  Hardware Oriented Security and Trust (HOST)}, 2019, pp. 181--190.

\bibitem{8942134}
H.~Chen, C.~Fu, J.~Zhao, and F.~Koushanfar, ``{Gen{U}nlock: An Automated
  Genetic Algorithm Framework for Unlocking Logic Encryption},'' in \emph{2019
  IEEE/ACM International Conference on Computer-Aided Design (ICCAD)}, 2019,
  pp. 1--8.

\bibitem{azar2020nngsat}
K.~Z. Azar, H.~M. Kamali, H.~Homayoun, and A.~Sasan, ``{NNgSAT: Neural Network
  guided SAT Attack on Logic Locked Complex Structures},'' in \emph{2020
  IEEE/ACM International Conference On Computer Aided Design (ICCAD)}, 2020.

\bibitem{sisejkovic2021logic}
D.~Sisejkovic, L.~M. Reimann, E.~Moussavi, F.~Merchant, and R.~Leupers,
  ``{Logic Locking at the Frontiers of Machine Learning: A Survey on
  Developments and Opportunities},'' \emph{arXiv}, 2021.

\bibitem{9006720}
A.~Alaql, D.~Forte, and S.~Bhunia, ``{Sweep to the Secret: A Constant
  Propagation Attack on Logic Locking},'' in \emph{2019 Asian Hardware Oriented
  Security and Trust Symposium (AsianHOST)}, 2019, pp. 1--6.

\bibitem{RANE2021Roshanisefat}
\BIBentryALTinterwordspacing
S.~Roshanisefat, H.~Mardani~Kamali, H.~Homayoun, and A.~Sasan, ``{RANE}: An
  open-source formal de-obfuscation attack for reverse engineering of logic
  encrypted circuits,'' ser. GLSVLSI '21.\hskip 1em plus 0.5em minus
  0.4em\relax New York, NY, USA: Association for Computing Machinery, 2021, p.
  221–228. [Online]. Available: \url{https://doi.org/10.1145/3453688.3461760}
\BIBentrySTDinterwordspacing

\bibitem{Alrahis2022OMLA}
L.~Alrahis, S.~Patnaik, M.~Shafique, and O.~Sinanoglu, ``{OMLA}: An oracle-less
  machine learning-based attack on logic locking,'' \emph{IEEE Transactions on
  Circuits and Systems II: Express Briefs}, vol.~69, no.~3, pp. 1602--1606,
  2022.

\bibitem{9474039}
L.~Alrahis, S.~Patnaik, F.~Khalid, M.~A. Hanif, H.~Saleh, M.~Shafique, and
  O.~Sinanoglu, ``{{GNNU}nlock: Graph Neural Networks-based Oracle-less
  Unlocking Scheme for Provably Secure Logic Locking},'' in \emph{2021 Design
  Automation and Test in Europe Conf. Exhibition (DATE)}, 2021.

\bibitem{sisejkovic2021challenging}
D.~Sisejkovic, F.~Merchant, L.~M. Reimann, H.~Srivastava, A.~Hallawa, and
  R.~Leupers, ``{Challenging the Security of Logic Locking Schemes in the Era
  of Deep Learning: A Neuroevolutionary Approach},'' \emph{ACM Journal on
  Emerging Technologies in Computing Systems (JETC)}, 2021.

\bibitem{UNTANGLE2021Alrahis}
L.~Alrahis, S.~Patnaik, M.~A. Hanif, M.~Shafique, and O.~Sinanoglu,
  ``{UNTANGLE}: Unlocking routing and logic obfuscation using graph neural
  networks-based link prediction,'' in \emph{2021 IEEE/ACM International
  Conference On Computer Aided Design (ICCAD)}, 2021, pp. 1--9.

\bibitem{Shamsi2022MLattackLUTLL}
\BIBentryALTinterwordspacing
K.~Shamsi and G.~Zhao, ``An oracle-less machine-learning attack against
  lookup-table-based logic locking,'' in \emph{Proceedings of the Great Lakes
  Symposium on VLSI 2022}, ser. GLSVLSI '22.\hskip 1em plus 0.5em minus
  0.4em\relax New York, NY, USA: Association for Computing Machinery, 2022, p.
  133–137. [Online]. Available: \url{https://doi.org/10.1145/3526241.3530377}
\BIBentrySTDinterwordspacing

\bibitem{Swarup2021SCOPE}
A.~Alaql, M.~M. Rahman, and S.~Bhunia, ``{SCOPE}: Synthesis-based constant
  propagation attack on logic locking,'' \emph{IEEE Transactions on Very Large
  Scale Integration (VLSI) Systems}, vol.~29, no.~8, pp. 1529--1542, 2021.

\bibitem{Alrahis2022GNNUnlockplus}
L.~Alrahis, S.~Patnaik, M.~A. Hanif, H.~Saleh, M.~Shafique, and O.~Sinanoglu,
  ``{GNNU}nlock+: A systematic methodology for designing graph neural
  networks-based oracle-less unlocking schemes for provably secure logic
  locking,'' \emph{IEEE Transactions on Emerging Topics in Computing}, vol.~10,
  no.~3, pp. 1575--1592, 2022.

\bibitem{Alrahis2022MuxLink}
L.~Alrahis, S.~Patnaik, M.~Shafique, and O.~Sinanoglu, ``Mux{L}ink:
  Circumventing learning-resilient mux-locking using graph neural network-based
  link prediction,'' in \emph{2022 Design Automation \& Test in Europe
  Conference \& Exhibition (DATE)}, 2022, pp. 694--699.

\bibitem{8607163}
P.~Chakraborty, J.~Cruz, and S.~Bhunia, ``{{SAIL}: Machine Learning Guided
  Structural Analysis Attack on Hardware Obfuscation},'' in \emph{2018
  AsianHOST}, 2018, pp. 56--61.

\bibitem{9530566}
L.~Alrahis, A.~Sengupta, J.~Knechtel, S.~Patnaik, H.~Saleh, B.~Mohammad,
  M.~Al-Qutayri, and O.~Sinanoglu, ``{GNN-RE: Graph Neural Networks for Reverse
  Engineering of Gate-Level Netlists},'' \emph{IEEE Trans. on Computer-Aided
  Design of Integrated Circuits and Systems}, 2021.

\bibitem{subhajitreverse}
S.~D. Chowdhury, K.~Yang, and P.~Nuzzo, ``Re{IGNN}: State register
  identification using graph neural networks for circuit reverse engineering,''
  in \emph{2021 IEEE/ACM International Conference On Computer Aided Design
  (ICCAD)}, 2021, pp. 1--9.

\bibitem{9300256}
J.~Sweeney, V.~Mohammed~Zackriya, S.~Pagliarini, and L.~Pileggi, ``{Latch-Based
  Logic Locking},'' in \emph{2020 IEEE International Symposium on Hardware
  Oriented Security and Trust (HOST)}, 2020, pp. 132--141.

\bibitem{9702267}
Y.~Hu, Y.~Zhang, K.~Yang, D.~Chen, P.~A. Beerel, and P.~Nuzzo, ``Fun-{SAT}:
  Functional corruptibility-guided {SAT}-based attack on sequential logic
  encryption,'' in \emph{2021 IEEE International Symposium on Hardware Oriented
  Security and Trust (HOST)}, 2021, pp. 281--291.

\bibitem{kotary2021endtoend}
J.~Kotary, F.~Fioretto, P.~Van~Hentenryck, and B.~Wilder, ``{End-to-end
  constrained optimization learning: A survey},'' \emph{arXiv}, 2021.

\bibitem{georgila2009using}
K.~Georgila, ``{Using Integer Linear Programming for Detecting Speech
  Disfluencies},'' in \emph{2009 Annual Conference of the North American
  Chapter of the Association for Computational Linguistics}, 2009.

\bibitem{sisejkovic2021deceptive}
D.~Sisejkovic, F.~Merchant, L.~M. Reimann, and R.~Leupers, ``{Deceptive Logic
  Locking for Hardware Integrity Protection against Machine Learning
  Attacks},'' \emph{IEEE Transactions on Computer-Aided Design of Integrated
  Circuits and Systems}, 2021.

\bibitem{6616532}
J.~Rajendran, H.~Zhang, C.~Zhang, G.~S. Rose, Y.~Pino, O.~Sinanoglu, and
  R.~Karri, ``Fault analysis-based logic encryption,'' \emph{IEEE Transactions
  on Computers}, vol.~64, no.~2, pp. 410--424, 2015.

\bibitem{mertensat}
M.~Merten, M.~E. Djeridane, S.~Huhn, and R.~Drechsler, ``{SAT}-based key
  determination attack for improving the quality assessment of logic locking
  mechanisms.''

\end{thebibliography}
\end{document}